\documentclass[%
reprint,
superscriptaddress,
amsmath,amssymb,
aps,
prl,
]{revtex4-2}

\usepackage{xcolor}
\newif\ifmarked \markedfalse
\newcommand{\rev}[1]{\ifmarked\textcolor{red}{#1}\else#1\fi}

\usepackage{graphicx}
\usepackage{dcolumn}
\usepackage{bm}
\usepackage{xcolor}

\begin{document}

\title{Programmable Synthetic Motion at a Time-Varying Interface}

\author{A. C. Harwood}
\altaffiliation{These authors contributed equally to this work}
\email{a.harwood22@imperial.ac.uk}
\affiliation{Blackett Laboratory, Department of Physics, Imperial College London, London SW7 2AZ, UK}

\author{D. Cielecki}
\thanks{These authors contributed equally to this work.}
\affiliation{Blackett Laboratory, Department of Physics, Imperial College London, London SW7 2AZ, UK}

\author{T. V. Raziman}
\affiliation{Blackett Laboratory, Department of Physics, Imperial College London, London SW7 2AZ, UK}
\affiliation{Department of Mathematics, Imperial College London, London, UK}

\author{S. A. Maier}
\affiliation{Blackett Laboratory, Department of Physics, Imperial College London, London SW7 2AZ, UK}
\affiliation{School of Physics and Astronomy, Monash University, Clayton, VIC 3800, Australia}

\author{S. Vezzoli}
\affiliation{Blackett Laboratory, Department of Physics, Imperial College London, London SW7 2AZ, UK}

\author{R. Sapienza}
\email{r.sapienza@imperial.ac.uk}
\affiliation{Blackett Laboratory, Department of Physics, Imperial College London, London SW7 2AZ, UK}

\date{\today}

\begin{abstract}
Space-time metamaterials that exhibit synthetic motion promise arbitrary control over the momentum, frequency and energy of scattered light, but realising the required space-time modulation in a programmable way remains a challenge. Here we program synthetic motion using a single spatial light modulator in a 4f geometry which imprints a continuously tunable pulse-front tilt onto a high-intensity pump pulse, inducing reflectivity modulations at a sub-wavelength indium tin oxide thin film with synthetic velocities spanning the sub- and superluminal regimes. The angle-resolved spectrum of a scattered probe pulse reveals space-time diffraction patterns whose gradient and bandwidth vary continuously with synthetic velocity, in excellent agreement with theory. Splitting the shaped pump into two independently controlled pulses yields space-time double-slit diffraction with tunable fringe separation and frequency-momentum gradient. This programmable platform opens a path towards non-linear and periodic space-time trajectories for tabletop analogue studies of relativistic phenomena and space-time metasurfaces.

\end{abstract}

\maketitle

\section*{Introduction}

Spatial metasurfaces harness subwavelength structuring  to sculpt the momentum of light, enabling flat optics, beam steering, and holography~\cite{chenReviewMetasurfacesPhysics2016a}, yet leave the frequency spectrum unchanged due to their temporal invariance. Structuring the properties of a material on sub-cycle timescales instead breaks translational invariance in time and lifts the reciprocity constraints.  Time-varying metamaterials have emerged as a versatile platform for simultaneously manipulating the energy and spectral content of waves, beyond conventional metamaterials~\cite{galiffiPhotonicsTimevaryingMedia2022, shaltoutTimevaryingMetasurfacesLorentz2015a, lyubarovAmplifiedEmissionLasing2022a}. In the optical domain, this is enabled by photo-induced ultrafast modulations of subwavelength thin films of transparent conductive oxides (TCOs)~\cite{kinseyEpsilonnearzeroAldopedZnO2015, alamLargeOpticalNonlinearity2016, caspaniEnhancedNonlinearRefractive2016, liUltrafastSwitchingOptical2026}, leading to the observation of temporal refraction~\cite{karlFrequencyConversionTimevariant2020, zhouBroadbandFrequencyTranslation2020, brunoBroadFrequencyShift2020} and diffraction~\cite{tiroleDoubleslitTimeDiffraction2023, tiroleSaturableTimevaryingMirror2022}, as well as ultrafast control of gain and loss~\cite{galiffiOpticalCoherentPerfect2026} and polarisation~\cite{jaffrayAllopticalPolarizationControl2026}.

Spatiotemporal metamaterials that exhibit synthetic motion, a space-time structured modulation of the optical properties of the device~\cite{bolotovskiiVavilovCerenkovEffectDoppler1972}, combine the spectral control of time-varying media with the directional control of spatially structured media \cite{shaltoutSpatiotemporalLightControl2019a, liberalSpatiotemporalSymmetriesEnergymomentum2024}, and have been predicted to support a wealth of phenomena including amplification \cite{pendryGainTimedependentMedia2021, crombAmplificationWavesRotating2020a, lyubarovAmplifiedEmissionLasing2022a}, complex pulse shaping \cite{ostrovskiiBoundaryAcceleratingMotion1975, kinderScatteringChirpingAccelerated2025}, nonreciprocity \cite{sounasNonreciprocalPhotonicsBased2017}, plasmonic coupling \cite{razimanSurfacePlasmonPolariton2025}, analogue relativistic phenomena~\cite{koufidisCouplingLightWaves2025, bahramiElectrodynamicsAcceleratedModulationSpaceTime2023, horsleyQuantumElectrodynamicsTimevarying2023}, and controllable quantum emission~\cite{sloanControllingTwophotonEmission2022, kort-kampSpaceTimeQuantumMetasurfaces2021a}. These prospects are now accessible, with recent studies of space-time modulated systems demonstrating magnet-less, non-reciprocal scattering~\cite{guoNonreciprocalMetasurfaceSpace2019}, frequency-momentum splitting of light pulses~\cite{jaffraySpatiospectralOpticalFission2025, ballSpaceTimeKnifeEdgeEpsilonNearZero2024a}, ultrafast transient lensing via space-time refraction~\cite{fanUltrafastWavefrontShaping2023}, and superluminal Doppler effect~\cite{harwoodSpacetimeOpticalDiffraction2025}. 

\rev{Structured light provides precise control over the full space-time form of optical pulses, from space-time wave packets with arbitrary group velocities \cite{kondakciOpticalSpacetimeWave2019, yessenovSpacetimeWavePackets2022a} to spatiotemporal optical vortices carrying orbital angular momentum \cite{hancockFreespacePropagationSpatiotemporal2019, yaoOrbitalAngularMomentum2011} and optical skyrmions with topologically structured polarisation \cite{forbesStructuredLight2021, shenOpticalSkyrmionsOther2024a}. Structured laser pulses can in turn be used to modulate matter in space and time \cite{shaltoutSpatiotemporalLightControl2019a}, an approach pursued at relativistic and ultrafast extremes \cite{piccardoTrendsRelativisticLaser2025} that has enabled optical pulses with tuneable synthetic velocities in bulk laser-driven plasma wakefields \cite{libermanDirectObservationWakefield2025, froulaSpatiotemporalControlLaser2018, caizerguesPhaselockedLaserwakefieldElectron2020}, analogue Hawking radiation \cite{belgiornoHawkingRadiationUltrashort2010a}, synchrotron radiation \cite{henstridgeSynchrotronRadiationAccelerating2018a} and superradiant light sources \cite{vazSpacetimeBeamsTunable2026}. However, in bulk media the structured pulse must propagate over long distances, during which nonlinear effects distort its programmed space-time form, whereas designing space-time modulations at interfaces offers a compact experimental paradigm free from propagation effects.}

\begin{figure*}[t]
\centering
\includegraphics[width=\textwidth]{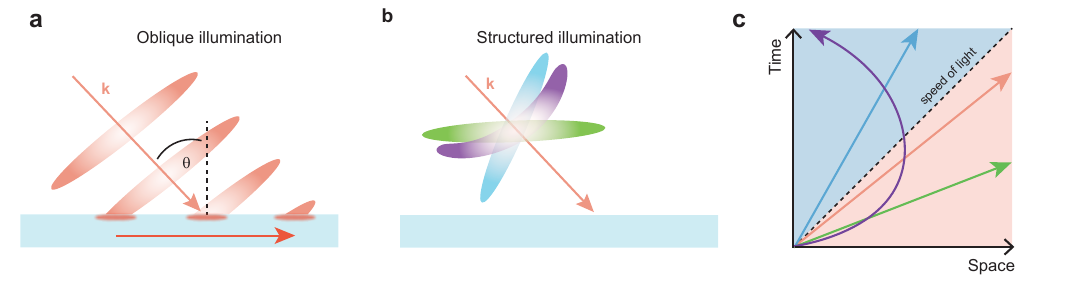}
\caption{\textbf{Synthetic motion at a time-varying interface.} (a) Oblique illumination of a pump pulse at angle $\theta$ illuminates a region that sweeps across the sample surface at a fixed synthetic velocity, illustrated at four snapshots in time, restricted to the superluminal regime. (b) Spatio-temporal structuring of the pump pulse decouples the synthetic velocity from the angle of incidence, giving access to the full space-time trajectory of the illumination pattern within a single optical setup. (c) The corresponding space-time trajectories accessible under structured illumination span subluminal and superluminal regimes \rev{and, in principle, accelerating paths}, in contrast to the fixed superluminal trajectory imposed by oblique illumination.}
\label{fig:1}
\end{figure*}

Here we demonstrate a platform for programmable space-time modulation at an indium tin oxide (ITO) interface. Using a single spatial light modulator (SLM) in a 4f pulse-shaping geometry, we achieve continuous, in-situ control over the synthetic velocity of reflectivity modulation, spanning subluminal to superluminal regimes. Angularly resolved space-time diffraction spectra of probe pulses scattered by the modulation exhibit a velocity-dependent momentum-frequency gradient. Throughout, experimental results are matched to numerical simulations based on operator theory~\cite{horsleyEigenpulsesDispersiveTimeVarying2023, harwoodSpacetimeOpticalDiffraction2025} and an analytical Fourier model, providing a predictive description of programmable synthetic motion. We further realise a space-time double slit with tuneable velocity and space-time separation, demonstrating independent control over the gradient and spacing of the diffraction fringes.

\section*{Generation of programmable synthetic motion}

A laser pulse incident obliquely at an angle $\theta$ on a surface illuminates a spot that traverses a space-time trajectory with velocity $v = c/\sin(\theta)$ (Fig.~\ref{fig:1}a), modulating the material properties of the surface along its path. Such synthetic motion transports no mass or information and can travel at speeds equal to and greater than the speed of light. The angle of incidence fixes the speed, however, restricting this scheme to superluminal linear motion and requiring mechanical realignment for any adjustment. Tilting the front of an ultrafast pulse so that it arrives at different locations at different times (Fig.~\ref{fig:1}b)~\cite{akturkPulsefrontTiltCaused2004} decouples the synthetic velocity from the angle of incidence, making subluminal and superluminal regimes continuously accessible without mechanical adjustment. {\rev{Full spatiotemporal control of the pulse extends this in principle to non-uniform trajectories such as acceleration~\cite{yessenovAcceleratingDeceleratingSpaceTime2020}, which we discuss as an outlook (Fig.~\ref{fig:1}c).}

We realise spatio-temporal pulse shaping by imprinting a tunable phase profile onto the joint frequency-momentum spectrum of the 180\,fs pulses, centred at 1300\,nm/230\,THz, using a 2D space-time pulse shaper (Fig.~\ref{fig:2}a, inset) \cite{kondakciOpticalSpacetimeWave2019}. The pulse shaper comprises a reflective grating, a cylindrical lens and a reflective SLM in a folded 4f configuration, placed at the back focal plane of a final spherical lens that focuses the pulse at the sample. The horizontal rulings of the grating disperse the frequency content of the incident pulses vertically, which are then collimated by the cylindrical lens, leaving the horizontal momentum unchanged. With the SLM positioned at the joint Fourier plane of the grating and focussing lens, its two axes map directly onto the frequency and transverse momentum of the pulse, such that any phase pattern imprinted on the SLM structures the spatio-temporal form of the pulse at the sample plane.

By imprinting a quadratic phase pattern -- which behaves as a cylindrical Fresnel lens -- either along the frequency axis or the momentum axis of the SLM, we can stretch the pulse in time or space at the sample plane.
Rotating this pattern by angle $\xi_\text{SLM}$ (Fig.~\ref{fig:2}a) relative to the frequency axis couples the spatial and temporal degrees of freedom, enabling arbitrary rotation of the pulse-front. However, the synthetic velocity of the pulse at the sample is not determined by the pulse-front tilt alone but also by the finite sizes of the pulse in space and time, and its projection onto the obliquely orientated sample surface. For a temporally narrow pulse with no front tilt, oblique incidence at angle $\theta$ already imparts a synthetic velocity $v = c/\sin(\theta)$ to the illuminated spot, and the pulse-front tilt introduced by $\xi_\text{SLM}$ acts to modify this baseline (see Supplemental Fig.~S8).

\begin{figure*}[t]
\centering
\includegraphics[width=\textwidth]{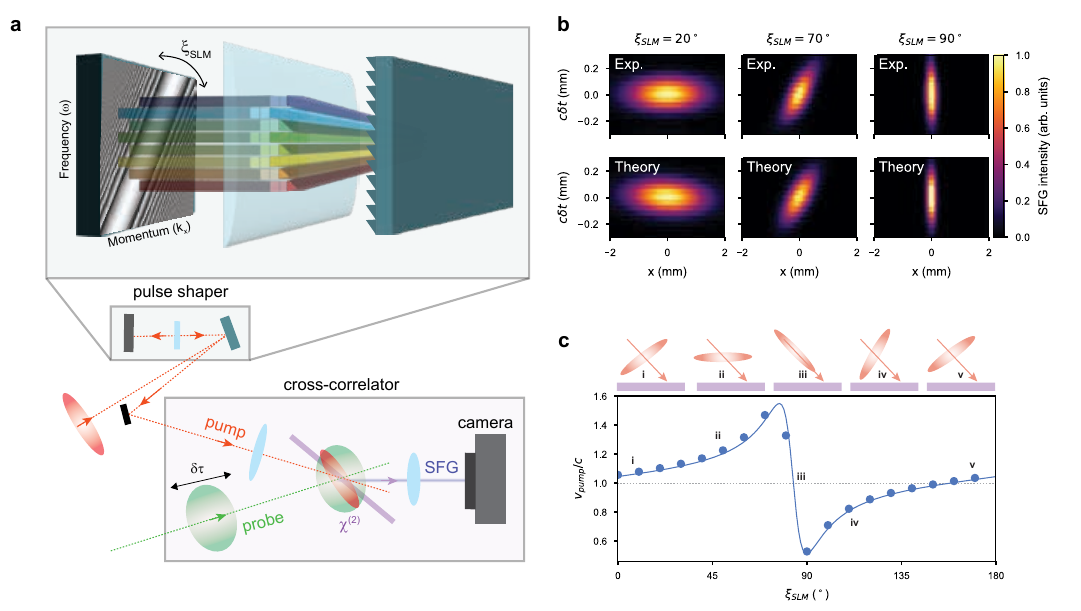}
\caption{\textbf{Shaping and characterising pump pulses with a programmable pulse front tilt.} (a) A pulse shaping setup, comprised of an SLM, cylindrical lens and grating in a 4f configuration, applies phase to the frequency and transverse momentum of the pump pulse. We characterise the shaped pulses with a spatially resolved cross-correlator setup based on sum-frequency generation (SFG) with a reference probe pulse in a $\chi^{(2)}$ nonlinear crystal (400\,$\mu$m GaP wafer). SFG signal recorded by a camera as a function of probe delay $\delta\tau$ yields a spatio-temporal map of the pump intensity profile. (b) Experimental (top) and simulated (bottom) SFG space-time maps $I_\text{SFG}(x, \delta\tau)$ for different rotation angles of frequency-momentum phase pattern $\xi_\text{SLM}$, demonstrating programmable rotation of the pulse front. (c) Synthetic velocity $v_\text{pump}$ can be subluminal or superluminal depending on the pulse front tilt (illustrations i--v), tuned by $\xi_\text{SLM}$. The synthetic velocity at the GaP sample extracted from SFG signals agrees strongly with analytical Fourier-based theory (solid lines).}
\label{fig:2}
\end{figure*}

\section*{Characterisation of synthetic motion}

We characterise the synthetic velocity of the structured pump pulse at the sample plane using a spatiotemporally resolved SFG cross-correlation scheme (Fig.~\ref{fig:2}a) \cite{potenzaThreeDimensionalImaging2004}. The shaped pump is focused onto a 400\,$\mu$m GaP wafer at an angle of $73^\circ$, where it overlaps with a probe pulse incident at $65^\circ$. Because of the lack of phase-matching conditions in GaP, frequency mixing only happens over the coherence length (a few $\mu$m), making the interaction length much shorter than the extension of the two pulses. The pump/probe angles match the geometry of the subsequent ITO diffraction experiments, corresponding to the Berreman resonance of the ITO nano-layer~\cite{tiroleSaturableTimevaryingMirror2022}. The probe pulse is stretched spatially to approximately 1000\,$\mu$m on the sample, three times the spatial extent of the pump beam, and stretched temporally to $\sim$640\,fs using a top-hat spectral filter (see Supplemental section ``Optical set-up''). The pump and probe durations and geometry in the cross-correlation experiment are chosen to match the subsequent diffraction experiments, ensuring experimental consistency between the two measurements.

Using a 2f imaging system, the SFG intensity profile is recorded on a camera for different values of probe delay $\delta\tau$. We combine these intensity profiles to build up a spatio-temporal map $I_\text{SFG}(x, \delta\tau)$ that encodes the overlap between the transverse trajectories of the probe and shaped pump pulse front at the sample. The spatio-temporal maps are simultaneously fitted to retrieve the spatial extents and pulse durations for both probe and unshaped pump, which we then use for theoretical modelling. Figure~2b shows experimental and simulated SFG space-time maps for $\xi_\text{SLM} = 20^\circ$, $70^\circ$ and $90^\circ$. Knowledge of the probe pulse parameters allows the retrieval of the trajectory of the shaped pump at the sample plane from the SFG map (see Supplemental section ``Sum frequency generation''), from which the synthetic velocity can be calculated.

\begin{figure*}[t]
\centering
\includegraphics[width=\textwidth]{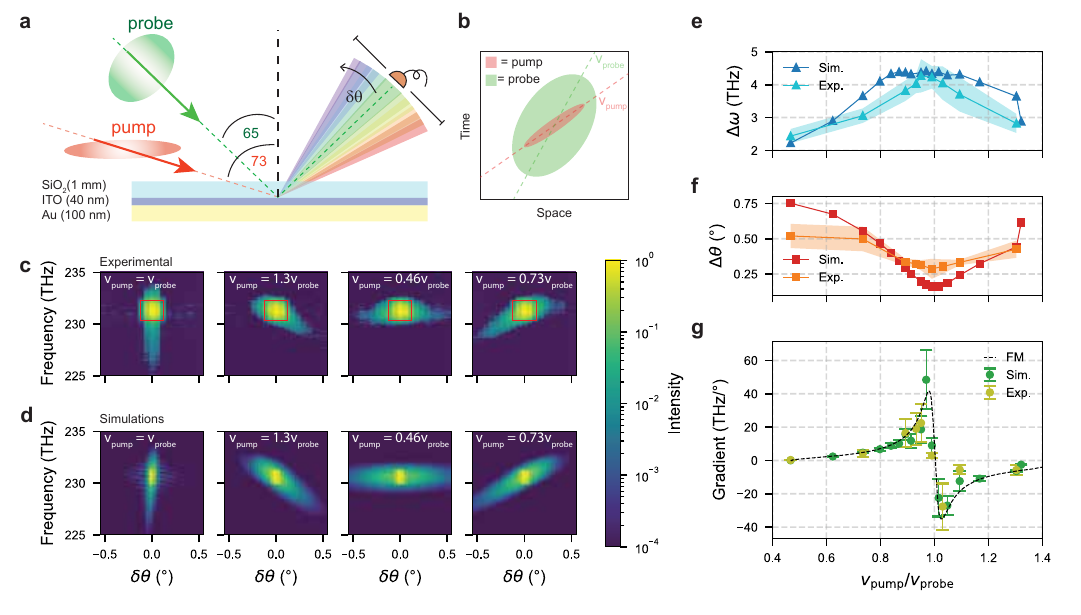}
\caption{\textbf{Space-time diffraction from programmable synthetic motion.} (a) A pump pulse with a programmable pulse-front tilt is incident on ITO, creating a synthetically moving transient modulation. (b) Keeping the phase velocity of the probe ($v_\text{probe}$) constant, we record the diffracted space-time spectrum as a function of angle from the reflected signal as we tune the synthetic velocity of the pump, $v_\text{pump}$. (c) Experimental and (d) simulated momentum-frequency hyperspectra of space-time diffraction at $v_\text{pump} = 1,\,1.3,\,0.46,\,0.73\, v_\text{probe}$, plotted on a logarithmic intensity scale. The extent of the unmodulated probe is marked by the red rectangle. (e) Spectral width $\Delta\omega$, (f) angular width $\Delta\theta$, defined by the full width at 1/1000th of the maximum, and (g) gradient of the diffraction pattern, extracted from experimental and simulated hyperspectra as a function of the pump-probe velocity ratio, in agreement with the analytical solution taken from our Fourier model (FM, dashed line). The gradient changes sign when velocities are equal, providing a direct signature of the transition between subluminal and superluminal regimes. \rev{Error bars denote the uncertainty in the gradient from a linear fit of the frequency of maximum spectral intensity as a function of $\delta\theta$ for each diffraction pattern.}}
\label{fig:3}
\end{figure*}

The observed synthetic velocity of the pump pulse, extracted from experiments on the angled sample, is plotted as a function of $\xi_\text{SLM}$ in Fig.~\ref{fig:2}c. Without any programmed front tilt, the pump pulse at $73^\circ$ incidence carries a baseline synthetic velocity of $1.05\, c$, while the probe pulse at $65^\circ$ carries a baseline of $1.10\,c$. Since the pulse-front tilt must first overcome this geometric contribution, the velocity curve is asymmetric about $\xi_\text{SLM} = 90^\circ$, with the luminal crossing occurring at $70^\circ$. The corresponding pulse orientations for different pattern rotations are sketched in Fig.~\ref{fig:2}c, i--v. The extracted velocities are in strong agreement with analytical theory (solid line, Fig.~\ref{fig:2}c) and also for Fresnel lenses of other parabolic strengths (Supplemental Fig. S6).

\section*{Programmable space-time diffraction}

The programmable synthetic velocity is now applied to drive space-time diffraction at a time-varying interface. The GaP wafer from the cross-correlator setup is replaced with an ITO sample comprising a 40\,nm ITO film ($\lambda_\text{ENZ} = 1320\,\text{nm}/227\,\text{THz}$) sandwiched between a 100\,nm gold layer and a SiO$_2$ substrate \cite{tiroleSaturableTimevaryingMirror2022}, Fig.~\ref{fig:3}a. The gold layer enhances the coupling of both pump and probe to the ITO film through the Berreman resonance of the bilayer, which occurs at approximately $65^\circ$. Crucially, decoupling the synthetic velocity from the pump angle of incidence means the pump can be held fixed close to this Berreman resonant angle across all programmed velocities, maximizing the induced refractive index modulation throughout.

Under intense optical excitation our sample undergoes a near-unity modulation of its refractive index, driven by intraband photocarrier excitation that results in a decrease in the plasma frequency of the sample due to a rapid increase in the effective mass of conduction band electrons before a slower return to equilibrium \cite{kinseyEpsilonnearzeroAldopedZnO2015, alamLargeOpticalNonlinearity2016, gurungControlUltrafastHot2025}. The probe experiences a subsequent ultrafast modulation of the sample's complex reflectivity coefficient $r = \rho\,e^{i\phi(t)}$, with $R = |r|^2$ changing from a few percent to $\sim$70\% \cite{tiroleSaturableTimevaryingMirror2022}. The amplitude modulation ($\rho$) broadens the diffracted probe spectrum, while the asymmetric temporal profile of $\phi(t)$, fast rise and slow decay, imparts first a larger red-shift and then a smaller blue shift as a function of delay. By using our 640\,fs probe pulse, we capture both dynamics to observe maximal spectral extension.

By using tilted pulses that exhibit synthetic motion across the surface of our sample, the pump-induced modulation creates a transient reflectivity aperture $r(x,t)$ that moves across the sample surface with programmable synthetic velocity $v_\text{pump}$ (Fig.~\ref{fig:3}b). The spatiotemporally larger probe pulse encapsulates the induced modulation\rev{, so that the diffracted light encodes the full space-time trajectory of the pump. The portion of the probe overlapping the modulation} is space-time diffracted, acquiring new frequencies at new angles, which we record by measuring the spectrum of the scattered light as a function of angle away from the reflected probe. \rev{The remaining portion of the probe that does not overlap the modulation is reflected without change in frequency or angle and remains within the unmodulated bandwidth}. Fig.~3c shows a selection of the experimental momentum-frequency hyperspectra of the diffracted probe, acquired for $v_\text{pump}/v_\text{probe} = 1,\,1.3,\,0.46$ and $0.73$, respectively. In each case, the diffraction pattern takes the form of an elongated distribution in the $(f, \theta)$ plane, with a momentum and frequency bandwidth that extends far beyond the bandwidths of the unmodulated probe pulse, illustrated by the red rectangle. With $v_\text{pump}$ varied as characterised in Fig.~\ref{fig:2}, the gradient and the bandwidth of the diffraction patterns change continuously. 

Experimental results are compared with numerical simulations of space-time diffraction, treating ITO as a Drude material whose plasma frequency $\omega_p(t)$ undergoes a sharp 10~fs drop and a slower 210~fs relaxation, with a maximum change of -10\%~\cite{tiroleSaturableTimevaryingMirror2022}.
We then simulate space-time diffraction from this system using an operator theory~\cite{horsleyEigenpulsesDispersiveTimeVarying2023} (see Methods for details) and extract three key observables: the spectral width $\Delta\omega$ (Fig.~\ref{fig:3}e), the angular width $\Delta\theta$ (Fig.~\ref{fig:3}f), and the gradient of both the simulated and experimental diffraction patterns (Fig.~\ref{fig:3}g). The spectral and angular widths reflect the spatio-temporal extent of the pump aperture and vary smoothly with $\xi_\text{SLM}$, consistent with the pulse profiles characterised in Fig.~\ref{fig:2}. Despite the trends in $\Delta\omega$ and $\Delta\theta$ being broadly consistent with theoretical predictions the match is complicated by uncertainty in the space-time profile of the modulation, see discussion for more details.

The gradient of the diffraction pattern (Fig.~\ref{fig:3}g) is less sensitive to these experimental uncertainties, and as a result the gradients extracted from the experimental data are in excellent agreement with the simulations based on operator theory (green) and the Fourier-based analytical theory (black) (see SI) across the full range of $\xi_\text{SLM}$. Notably, the gradient undergoes a sign change at $v_\text{pump} = v_\text{probe}$, serving as a direct experimental signature of the transition between subluminal and superluminal regimes and confirming programmable, continuous control over space-time diffraction at an ENZ interface. The range of accessible gradients that can be programmed \rev{continuously} is restricted to approximately $\pm40$\,THz/$^\circ$. The elliptical nature of the shaped pump, with widths that vary with $\xi_\text{SLM}$, limits the effective velocity of the space-time trajectory on the oblique film compared to a narrow beam.

\section*{Discrete synthetic motion}

To increase the complexity of the momentum-frequency synthesis of scattered light, we extend our experimental platform to modulation patterns that exhibit discrete synthetic motion\rev{, in which the modulation is composed of localised events separated in space and time, between which a discrete synthetic velocity can be attributed ~\cite{harwoodSpacetimeOpticalDiffraction2025}}. We generate a programmable space-time double slit by replacing the rotated Fresnel lens SLM phase pattern with a rotated periodic phase pattern (see Supplemental section ``Double slit''). This splits the pump into a pair of spatially and temporally separated pulses (Fig.~\ref{fig:4}a), each with an approximate fluence of 36\,mJ/cm$^2$. When used to illuminate the sample, \rev{the two pulses generate compact,} localised reflectivity modulations with spatial and temporal separation $\Delta x$ and $\Delta t$, \rev{whose relative kinematics define} a discrete synthetic velocity $v = \Delta x/\Delta t$. The programmable pulse shaper enables accurate and independent control of the modulation velocity while keeping their spatial separation constant (Fig.~\ref{fig:4}b), or of their space-time separation $d_s$ while keeping their velocity constant (Fig.~\ref{fig:4}c).

\begin{figure*}[t]
\centering
\includegraphics[width=\textwidth]{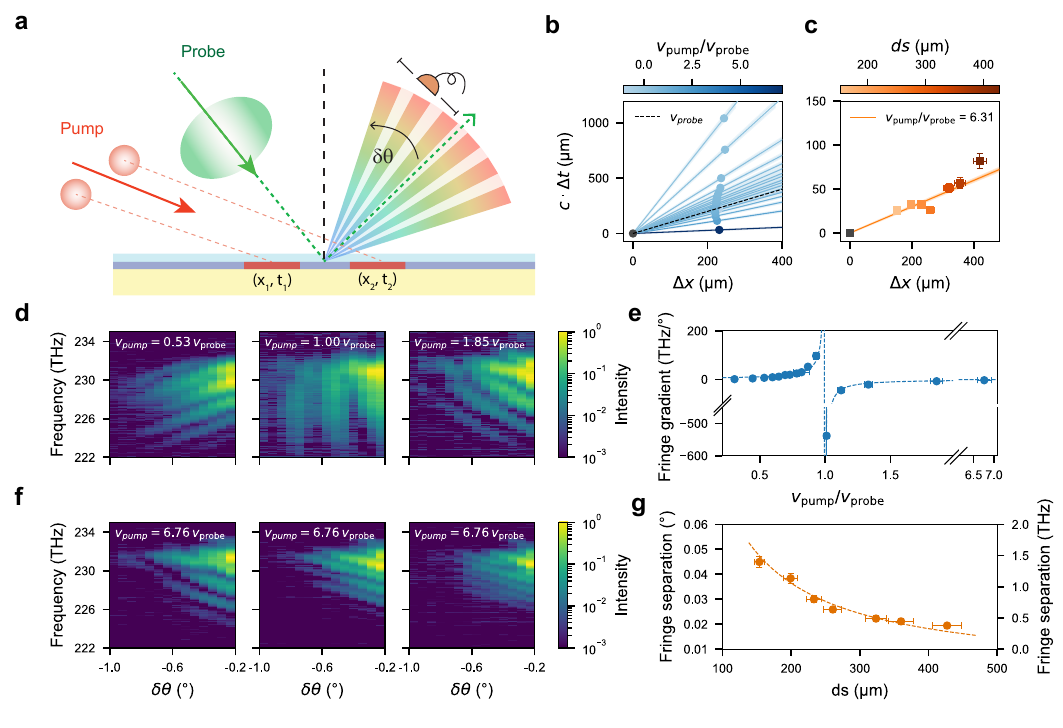}
\caption{\textbf{Programmable space-time double slit.} (a) The pump pulse is split into two pulses that generate modulated apertures with programmable space-time separation and discrete velocity. Space-time diffraction patterns are detected by scanning away from the reflected signal to improve visibility. While keeping one pulse fixed (black at origin), we program (b) the discrete synthetic velocity for a constant spatial separation as well as (c) the space-time separation $d_s$ for a constant velocity. The acquired diffraction patterns exhibit interference fringes in momentum-frequency space \rev{that depend upon (d, e) the discrete synthetic velocity and (f, g) the space-time separation between the modulations (ds), in agreement with the full analytical model (dashed lines)}.}

\label{fig:4}
\end{figure*}

Momentum-frequency hyperspectra are acquired for a range of angles away from the directly reflected probe signal to isolate the lower intensity diffraction from the undiffracted light, improving the signal-to-noise ratio. First, we study the dependence of the diffraction pattern upon the discrete synthetic velocity of the two modulated spots, with three examples plotted in Fig.~\ref{fig:4}d for velocities $v_\text{pump} = 0.53\,v_\text{probe}$, $1.00\,v_\text{probe}$ and $1.85\,v_\text{probe}$. The double slit diffraction pattern is signified by clear, diagonal interference fringes, which extend well beyond the input probe bandwidth of $\sim$2\,THz. As with the gradient of the diffraction patterns from continuous motion explored in Fig.~\ref{fig:3}, the gradient of the fringes depends on the ratio $v_\text{pump}/v_\text{probe}$. \rev{Approximating the two apertures as points, the gradient of the fringes follows from the ratio of their periods in frequency and angle (see Supplemental section ``Fringes of space-time double slit''),}
\begin{equation}
    {\frac{\delta f}{\delta\theta} = \frac{f_0\,\beta\cos\theta_0}{1-\beta\sin\theta_0},}
    \label{eq:gradient}
\end{equation}
\rev{where $f_0$ is the central probe frequency and $\beta\sin\theta_0 = v_\text{pump}/v_\text{probe}$.} Here, unlike in Fig.~\ref{fig:3}, the gradient diverges at the luminal crossing \rev{where $v_\text{pump} = v_\text{probe}$}, so the gradients of the experimental diffraction patterns (Fig.~\ref{fig:4}e) extend far beyond the $\pm40$\,THz/$^\circ$ range demonstrated with continuous motion\rev{, in agreement with Eq.~(\ref{eq:gradient})}.

To further explore the programmability, we study the dependence of the fringe separation on the space-time separation of the two modulations by keeping the velocity fixed at $v_\text{pump} = 6.31\,v_\text{probe}$ (Fig.~\ref{fig:4}c, orange). The hyperspectra demonstrate visibly constant gradients as the space-time separation increases (Fig.~\ref{fig:4}f), with decreasing frequency-momentum fringe separation, extending the space- and time-only \cite{tiroleDoubleslitTimeDiffraction2023} double slit diffraction experiments where only fringe separation in momentum or frequency can be controlled. Fig.~\ref{fig:4}f plots the dependence of the fringe separation in THz and degrees as a function of the space-time separation, showing excellent agreement with the inverse relationship expected from theory (see Supplemental section ``Fringes of space-time double slit'').

\section*{Discussion}

\rev{Programmable space-time pulse shaping decouples synthetic motion from the illumination geometry, turning the trajectory of the modulation into an experimental parameter that can be explored systematically. This extends past work beyond a narrow range of superluminal velocities into the previously inaccessible subluminal regime, while the pump remains at the resonant angle of the sample for maximal coupling and modulation efficiency. Space-time diffraction can therefore be studied continuously across the luminal crossing, $v_\text{pump} \approx v_\text{probe}$, where the measured gradients follow the Fourier-based theory through the transition from relativistic to super-relativistic Doppler scattering, a transition at the heart of theoretical proposals for amplification and compression~\cite{pendryGainTimedependentMedia2021, horsleyTravelingWaveAmplification2024} and of analogue studies of Hawking radiation~\cite{horsleyQuantumElectrodynamicsTimevarying2023}, gravitational waves~\cite{koufidisGravitationalWavesOptical2025, koufidisCouplingLightWaves2025} and \v{C}erenkov radiation~\cite{bolotovskiiVavilovCerenkovEffectDoppler1972, sloanControllingTwophotonEmission2022}. Discrete synthetic motion extends the platform to frequency-momentum synthesis of light, as a velocity defined by the slope between localised events is independent of their separation, so the orientation and spacing of the fringes can be programmed independently within the modulation bandwidth, which was not possible in previous experiments~\cite{harwoodSpacetimeOpticalDiffraction2025}.}

\rev{By generating synthetic motion across the subluminal and superluminal regimes, the synthesised hyperspectrum is rotated throughout momentum-frequency space, providing reconfigurable ultrafast beam steering and spectral translation~\cite{shaltoutSpatiotemporalLightControl2019a}. We model this synthesis analytically and numerically from the space-time structure of the pump retrieved by spatially resolved cross-correlation, with the numerical model also incorporating the photoexcitation of the medium. Both reproduce the gradient and period of the diffraction patterns across all programmed velocities and separations, limited only near the luminal crossing by the uncertainty in the linear fit to the diffraction maximum where the pattern becomes near-vertical. The bandwidth, which additionally depends on the precise modulation profile, reaches $\approx$10~THz, slightly beyond our numerical model, which we attribute to the perturbative approximation of a modulation that becomes non-perturbative at high intensities~\cite{tiroleSaturableTimevaryingMirror2022, reshefPerturbativeDescriptionNonlinear2017}, two-beam coupling~\cite{paulTwobeamCouplingHot2021} and the finite resolution of our hyperspectral detection. Broader scattering could be achieved with shorter pump pulses~\cite{lustigTimerefractionOpticsSingle2023, liUltrafastSwitchingOptical2026} or higher fluences~\cite{harwoodCompetingAugerlikeIntraband2026}, while materials with a more symmetric temporal response could balance the bandwidth about the probe frequency.}

\rev{Beyond the continuous and discrete motion demonstrated here, the platform could be extended to more complex trajectories or to larger ensembles of pump pulses, enabling photonic space-time crystals~\cite{sharabiSpatiotemporalPhotonicCrystals2022}, disordered space-time media~\cite{sharabiDisorderedPhotonicTime2021, coppolaroExploringAperiodicOrder2025} and higher-dimensional modulation~\cite{enghetaFourdimensionalOpticsUsing2023a}, within two main limits. The first is intrinsic to the medium, as driving the ITO into its saturable regime requires $\sim$100~GW/cm$^2$, so pulse arrays spread over a larger space-time volume and further attenuated by the $\sim$35\% shaper efficiency struggle to modulate it efficiently. The decay of the modulation over hundreds of femtoseconds also sets how closely features can be spaced, while repeated modulation of the same region pushes the accumulated fluence towards the damage threshold~\cite{hayranokDispersion2022, harwoodCompetingAugerlikeIntraband2026}. The second arises from the pulse shaper, as the phase difference between adjacent SLM pixels is limited to $\pm\pi$ without aliasing, so the finite pixel sampling bounds the maximum space-time extent of the shaped pattern, while phase jumps and gaps between pixels scatter energy into unwanted diffraction orders, further reducing the efficiency as this limit is approached.}

\section*{Conclusion}

\rev{We have demonstrated programmable synthetic motion at a time-varying interface, tuning the synthetic velocity continuously from the subluminal to the superluminal regime without mechanical realignment for both single and double slit space-time modulations. Programming the continuous synthetic velocity of a single aperture controls the gradient of the diffraction pattern, including its sign reversal at the luminal crossing, while independent programming of the discrete velocity and space-time separation of a double slit yields orthogonal control over the gradient and spacing of its fringes, both in good agreement with analytical Fourier theory and operator-based numerical simulations. Looking forward, combining the hyperspectral synthesis demonstrated here with ultrafast polarisation control~\cite{jaffrayAllopticalPolarizationControl2026} and programmable amplification or absorption~\cite{galiffiOpticalCoherentPerfect2026}, both realised in similar TCO-based samples, would define a high-dimensional active pulse shaper. The platform also makes reconfigurable tabletop analogues of relativistic phenomena accessible, and its programmable transluminal synthetic velocity applies directly to \v{C}erenkov radiation~\cite{bolotovskiiVavilovCerenkovEffectDoppler1972, sloanControllingTwophotonEmission2022}, whose emission could be switched on and steered towards convenient detection angles by tuning $v_\text{pump}$ across the phase velocity of light in the surrounding medium. Finally, programmable space-time pump and probe pulses lend themselves directly to autonomous pump-probe spectroscopy, in which the space-time structure and evolution of modulations and carrier dynamics are characterised without moving parts, extending self-driving laboratories~\cite{desaiSelfdrivingLabDiscovers2025} into the space-time domain.}

\section*{Methods}

\subsection*{Sample}

The time-varying medium used in this work is a thin film of indium tin oxide (ITO) with low optical loss from Prazisions Glas \& Optik GmbH. The 40\,nm ITO film is deposited on a glass cover slip and covered by a 100\,nm layer of gold. The gold layer enhances the reflectivity modulation and shifts the epsilon-near-zero frequency from 227\,THz to 232\,THz (1300\,nm) due to the change in the refractive index of the surrounding medium. The same sample has been used in \cite{tiroleSaturableTimevaryingMirror2022}.

\subsection*{Pulse shaper}

We integrated a 4f pulse shaper with the SLM into the pump-probe setup (see SI). Ultrafast near-infrared pulses at 231\,THz with a duration of $\sim$180\,fs are generated by an ORPHEUS (Light Conversion) optical parametric amplifier from a solid-state PHAROS (Light Conversion) laser. The laser output is p-polarised and aligned with the nematic axis of the SLM using a half-wave plate after the OPA. Eighty percent of the power is directed to a reflective diffraction grating (600\,lines/mm, 1.25\,$\mu$m blaze), and then Fourier transformed in the $y$-direction at the SLM display by a cylindrical lens ($f = 13$\,mm). The SLM pixels imprint a phase on the spectral components of the pulse. The SLM reflects the beam back through the cylindrical lens and grating, where the frequency components are recombined. The pump is then focused onto the sample, which is located at the Fourier plane of the SLM display.

\subsection*{SLM patterns}

\subsubsection*{Uniform motion}

The SLM axes were mapped into angular and spatial frequencies. Tilting the pulse front is equivalent to a linear gradient in the added phase variation between adjacent frequency components, inducing a uniform delay along the direction of the gradient. In practice, this equates to a quadratic phase profile in one direction, which stretches the pulse along that direction and rotates the pulse front by an angle dependent on the rotation. A cylindrical lens profile,
\begin{equation}
\phi(k_x, \omega) = \frac{8\pi}{W_\text{SLM}^2}
\left(k_x \cos\xi_\text{SLM} - \omega \sin\xi_\text{SLM}\right)^2,
\end{equation}
was used to induce the appropriate phase variation, symmetrically about the direction of the wavevector, where $k_x$ are the momentum components along the $x$-axis of the SLM and $\omega$ is the frequency of the pump mapped to the $y$-axis of the SLM. The incoming pump beam was aligned to the centre of the rotated cylindrical lens, with a curvature defined by the central width $W_\text{SLM}$ (in pixels) and a rotation angle $\xi_\text{SLM}$ (see Supplemental section ``SLM patterns'').

\subsubsection*{Programmable double slit}

We use the Gerchberg-Saxton (G-S) algorithm~\cite{gerchbergPracticalAlgorithmDetermination1972} to generate the programmable space-time double slit.
The algorithm computes the 1-D phase function $\phi(k_x)$ that, when applied to the gaussian momentum spectrum of the unshaped pump $f (k_x)$, has a fourier transform with intensity equal to the target function $D_\Delta(x)$ consisting of two gaussians with separation $\Delta$.
The target gaussians have the same width as the unshaped pump pulse.
We start with a cylindrical Fresnel lens pattern for $\phi$.
The G-S algorithm does the two steps repeatedly:
\begin{enumerate}
    \item Compute the transform $F(x)$ of $f(k_x)\exp[i\phi(k_x)]$. Let $\psi(x)$ be the argument of $F$
    \item Compute the inverse transform $f'(k_x)$ of $\sqrt{D_\Delta(x)}\exp[i\psi(x)]$. Update $\phi(k_x)$ with the argument of $f'$
\end{enumerate}
When $|F|^2$ converges to $D_\Delta$, $\phi(k_x)$ gives the required phase pattern to create the double-slit. As $\phi$ is one-dimensional, it is applied on the SLM by rotating to achieve the required space-time rotation of the double slit, and with a uniform phase in the orthogonal direction. We speed up the algorithm by weighting the peaks of $D_\Delta$ in step 2 to compensate for the peaks of $F$ being too small or big. Large ($>\frac{\pi}{2}$) variations between adjacent pixels of $\phi$ are smoothened. G-S algorithm is run for 30 rounds to compute phase patterns.

\subsection*{SFG experiment and retrieval of shaped pump}

The spatially-resolved cross-correlator was built using a $\chi^{(2)}$ nonlinear medium in which sum-frequency generation occurs when the pump and probe are spatio-temporally overlapped. A 400\,$\mu$m gallium phosphide wafer was used at the sample plane, with the pump incident at $73^\circ$ and the probe at $65^\circ$, in order to characterise the pulse shape under conditions consistent with the diffraction experiments. A GaP wafer replaced the ITO sample without altering its orientation relative to the beams, maintaining optical consistency. The sample plane was imaged onto a camera sensor using a 75\,mm lens in a 2f--2f system. Images were collected for a sequence of time delays between pump and probe pulses, and the SFG signal was filtered and collected as shown in Supplemental Figure S5a. The average power of both beams was kept constant across the entire dataset.

Each SFG camera image was fit to an axis-parallel gaussian of the form
\begin{equation*}
    I(x, y) = A \exp\left\{-\frac12 \left[\left(\frac{x-x_0}{\sigma_x}\right)^2 + \left(\frac{y-y_0}{\sigma_y}\right)^2\right]\right\} + B
\end{equation*}
where $A$ is the amplitude, $x/y_0$ is the centre, $\sigma_{x/y}$ is the width, and $B$ is the background. 1-D gaussians of $(A, x_0, \sigma_x)$ for a given $(W_{SLM}, \xi_{SLM})$ pair were stacked for different values of pump-probe delay $\tau$ to obtain $I_{\mathrm{SFG}}(x, \tau)$ images as shown in Supplemental Figure 5b, and to calculate the \textit{spread} $G_{\mathrm{SFG}}(x, \tau)$ (see Supplemental section ``Theory''). $I_{\mathrm{SFG}}$ images for all $(W_{SLM}, \xi_{SLM})$ pairs were fit simultaneously using only the spatiotemporal widths of the probe and unshaped pump as fitting parameters. $I_{\mathrm{SFG}}$ and the probe parameters were then used to retrieve the trajectory of the shaped pump on the sample and compute its velocity as in Fig.~\ref{fig:2}b.

\subsection*{ITO experiment}

The pump and probe pulses were spatio-temporally overlapped at the sample, each incident at $73^\circ$, $65^\circ$ respectively. The reflection of the diffracted probe was spatially filtered in horizontal direction by a slit aperture at the Fourier plane of a lens ($f=100$ mm), and then collected by a fibre-coupled Princeton Instruments NIRvana spectrometer (see Supplemental section ``Optical set-up''). Hyperspectral scans were acquired by measuring a spectrum for a range of slit and delay stage positions. Analysis was performed for the hyperspectral maps where the modulation of the probe has the highest extent in frequency and momentum. 

\subsection*{Numerical simulation}
We model the permittivity of the ITO layer as a time-varying Drude material, 
\begin{equation*}
    \epsilon(\omega, t) = \epsilon_\infty - \frac{\omega_p(t)^2}{\omega(\omega + i\gamma)}\,.
\end{equation*}
with constant background permittivity $\epsilon_\infty$ and damping constant $\gamma$, and time-varying plasma frequency $\omega_p(t) = \omega_{p0} - \Delta f(t)$, where $\omega_{p0}$ is the unmodulated plasma frequency and $\Delta$ is the maximum change.
$f$ is the convolution of the pump intensity at the location with a temporal response kernel $\tau$ having a fast rise time $t_{rise}$ and slow decay time $t_{decay}$,
\begin{equation*}
    \tau(t-t') = \frac{1}{2} \left[1 + \tanh\left(\frac{t-t'-t_{rise}}{t_{rise}}\right)\right]e^{-\left(\frac{t-t'-t_{rise}}{t_{decay}}\right)}\,,
\end{equation*}
normalised to $\max(f)=1$.
Due to the non-perturbative nonlinearity of ITO~\cite{reshefPerturbativeDescriptionNonlinear2017}, we opt for a simplified model in which the maximum modulation $\Delta$ is set to 10\%~\cite{ tiroleSaturableTimevaryingMirror2022} for all $\xi_\text{SLM}$, despite the varying pump fluences.
We also neglect the less prominent modulation of $\gamma$ at low fluences~\cite{harwoodCompetingAugerlikeIntraband2026}. The parameters used in the simulations are $\epsilon_\infty$=3.9, $\omega_{p0}$=450.5 THz, $\gamma$=130 THz. $t_{rise}$=10 fs, $t_{decay}$=210 fs.

We simulate space-time diffraction from the film using an operator-based approach for wave propagation in spatiotemporally modulated media~\cite{horsleyEigenpulsesDispersiveTimeVarying2023, harwoodSpacetimeOpticalDiffraction2025}.
The permittivity is converted to an operator representation that acts on the frequency and in-plane momentum of electromagnetic fields.
Maxwell's equations are solved in the operator framework  to find the electromagnetic fields in the system.
A transfer matrix method is used to compute the generalised reflection and transmission coefficients from the ITO/Au bilayer that include frequency-momentum mixing.
Projecting the incident probe spectrum on these coefficients and transforming the in-plane momenta to angles using $k_x = \omega/c_0 \sin(\theta)$ yields the hyperspectra in Fig.~\ref{fig:3}c.   

\section*{Data availability}
The data that support the findings of this study are available from the corresponding author upon reasonable request.

\section*{Code availability}
The code used for the numerical simulations and data analysis in this study is available from the corresponding author upon reasonable request.

\bibliographystyle{naturemag}
\bibliography{SLM_diffraction}

\subsection*{Acknowledgements}
The authors would like to thank Simon Horsley for useful discussions. This work was supported by The Val O’Donoghue Scholarship in Natural Sciences (ACH), the Engineering and Physical Sciences Research Council (EPSRC), grant number EP/Y015673 (TVR, SV, RS) and the Lee-Lucas Chair in Physics (SAM). We acknowledge computational resources and support provided by the Imperial College Research Computing Service (http://doi.org/10.14469/hpc/2232).

\subsection*{Author Contributions}
~\\
Conceptualisation: SV, RS\\
Methodology: ACH, DC, SV, TVR, RS\\
Software: TVR, ACH, DC\\
Investigation: ACH, DC, TVR\\
Visualisation: ACH, DC, RS \\
Writing - Original draft:  ACH\\
Writing - Reviewing and Editing: All authors \\
Supervision: RS, SV \\

\end{document}